\documentclass[reprint,
 amsmath,amssymb,
 aps,superscriptaddress
]{revtex4-2}

\usepackage{graphicx}% Include figure files
\usepackage{dcolumn}% Align table columns on decimal point
\usepackage{bm}% bold math
\usepackage{url}

\begin{document}

\preprint{}

\title{New Measurements of $^{71}$Ge Decay: Impact on the  Gallium Anomaly}% Force line breaks with \\

\author{J.I.\ Collar}
\email{collar@uchicago.edu}
\affiliation{Enrico Fermi Institute, Kavli Institute for Cosmological Physics, and Department of Physics\\
University of Chicago, Chicago, Illinois 60637, USA}

\affiliation{Donostia International Physics Center (DIPC), Paseo Manuel Lardizabal 4, 20018 Donostia-San Sebastian, Spain}

\author{S.G.\ Yoon}
\email{sgyoon@uchicago.edu}
\affiliation{Enrico Fermi Institute, Kavli Institute for Cosmological Physics, and Department of Physics\\
University of Chicago, Chicago, Illinois 60637, USA}

%\affiliation{%
%Enrico Fermi Institute, Kavli Institute for Cosmological Physics, and Department of Physics\\
%University of Chicago, Chicago, Illinois 60637, USA
%}


\date{\today}% It is always \today, today,

\begin{abstract}
A dedicated high-statistics measurement of the $^{71}$Ge half-life is found to be in accurate agreement with an accepted value of 11.43$\pm$0.03 d, eliminating a recently proposed route to bypass the ``gallium anomaly¨ affecting several neutrino experiments. Our data also severely constrain the possibility of $^{71}$Ge decay to low-energy excited levels of the $^{71}$Ga daughter nucleus as a solution to this puzzle. Additional unpublished measurements of this decay are discussed. Following the incorporation of this new information, the gallium anomaly survives with high statistical significance. 
\end{abstract}

\maketitle

%\tableofcontents

When exposed to intense radioisotopic neutrino sources ($^{51}$Cr and $^{37}$Ar) several gallium-based neutrino detectors (GALLEX \cite{gallex1,gallex2}, SAGE \cite{sage1,sage2}, BEST \cite{best1,best2})  display a $\sim$20\% deficit in the observed interaction rate with respect to the Standard Model  expectation.  This ``gallium anomaly" \cite{anomaly1,anomaly2} has been interpreted within the context of sterile neutrino oscillations \cite{anomaly1,steriles}. This perspective is nevertheless in high tension with other neutrino measurements, leading to an ongoing effort to find other possible explanations. Some involve new physics \cite{giunti,kopp,decoherence}, others concentrate on simpler scenarios where basic assumptions made in the interpretation of gallium experiments are closely examined. 

Two recent papers \cite{giunti,kopp} have pointed out that a slightly larger value of the half-life for the electron capture (EC) decay of $^{71}$Ge, in agreement with some of its individual measurements, can do away with the anomaly. The value required (T$_{1/2}\sim$12.5 d) is however not compatible with the latest adopted reference (T$_{1/2}=11.43\pm0.03$ d).  A $\sim$10\% branching ratio (BR) of this decay into an excited level(s) of the daughter $^{71}$Ga nucleus would accomplish a similar relaxation of   evidence for the anomaly \cite{kopp}. Both half-life and BR  affect the nuclear matrix element entering the calculation of the cross section for the relevant inverse process, $\sigma (\nu_{e} + ^{71}$Ga $\rightarrow$ e$^{-} + ^{71}$Ge) \cite{haxton}. This hypothetical excited level would have an energy below 232.4 keV (the Q value of $^{71}$Ge decay), perhaps complicating the observation of tell-tale de-excitation gammas \cite{kopp}. Other assumptions scrutinized in \cite{kopp} involve a correction to the BRs in the decay of  neutrino-emitting $^{51}$Cr sources employed by gallium experiments, as well as the impact that revised values of the $^{71}$Ge extraction efficiency would have for those.

\begin{figure}[!htbp]
\includegraphics[width=.8 \linewidth]{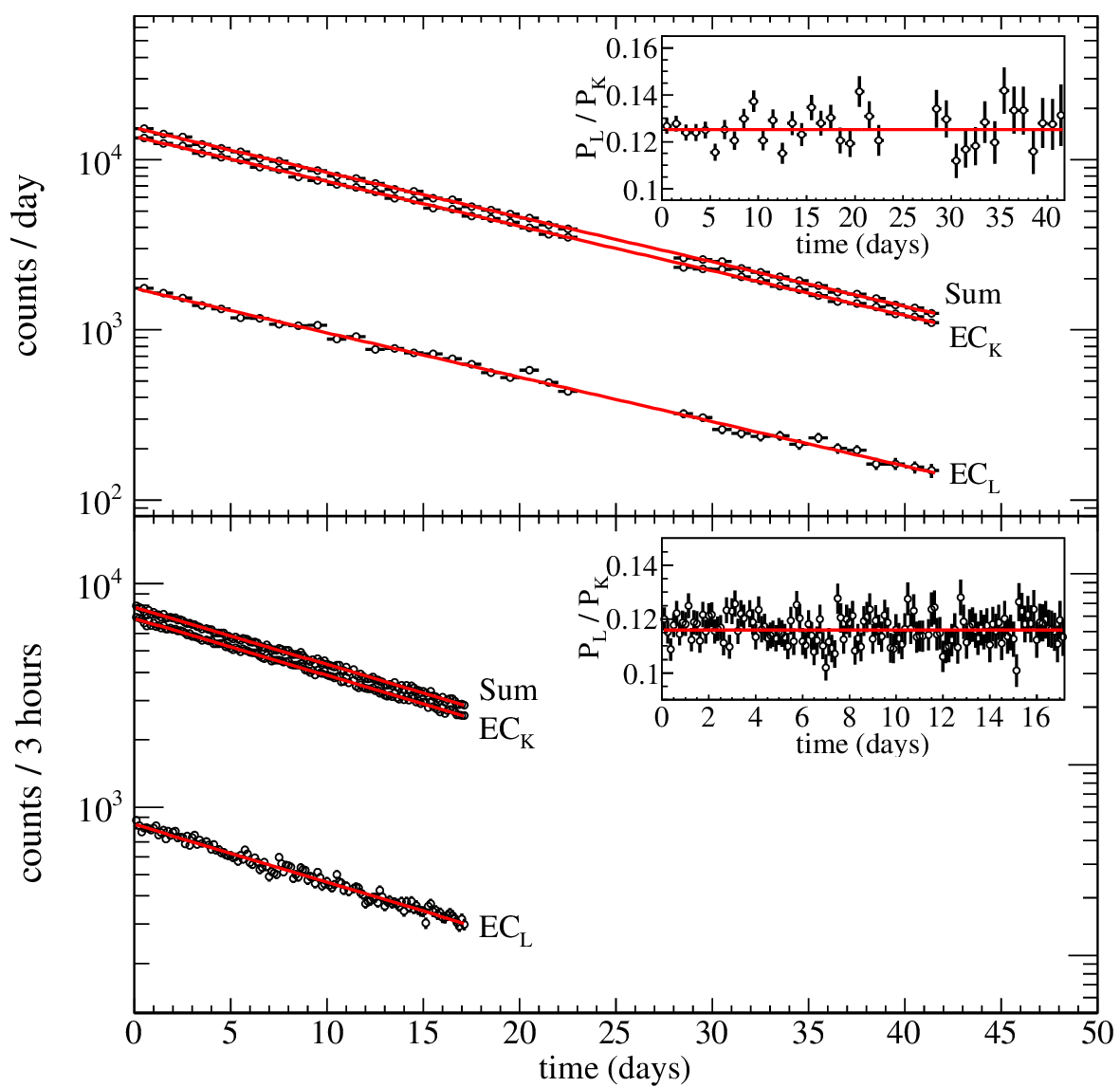}% Here is how to import EPS art
\caption{\label{fig:epsart} $^{71}$Ge decay rates  in the present  measurement (top) and the detector in \cite{ppc} (bottom). Vertical (statistical) error bars are encumbered by the data points.  Insets show the relative probability of EC from L and K shells. These include a small ($<\!\!<$1\%) correction for X-ray escape from each crystal. Red lines are exponential fits to the data (constant for insets).}
\end{figure}

In this brief note we describe a dedicated measurement tailored to test all aspects of the decay of $^{71}$Ge able to impact the interpretation of the gallium anomaly. A small (1 cm$^{3}$) n-type germanium diode \cite{qf} was used for this purpose. The device was initially shielded against environmental radiations using 10 cm of Pb in a laboratory benefiting from a 6 \frenchspacing{m.w.e.} overburden. A background spectrum was obtained over 2.7 days, following energy calibration using gamma emitters. The origins of all peaks visible in this spectrum (Fig.\ 2) are readily identifiable (neutron reactions, cosmogenic activations, U- and Th-chain radioimpurities, etc.). The detector was then activated in $^{71}$Ge via a four day exposure to a moderated $^{252}$Cf neutron source at the center of a 20 cm polyethylene sphere \cite{qf}. A production of approximately $5\times \!10^{5}$  $^{71}$Ge atoms was expected via simulation of the $^{70}$Ge neutron capture rate. The detector was returned to its shield. A total of 42 days of post-activation data were taken, with a single 5.7 day interruption due to failure of the data-acquisition system (Fig.\ 1). The system stored time-stamped individual event traces, allowing an arbitrary time binning. The energy spectrum spanned the range from a 0.5 keV threshold up to 250 keV, i.e., beyond the Q-value of $^{71}$Ge decay.

\begin{figure}[!htbp]
\includegraphics[width=.8 \linewidth]{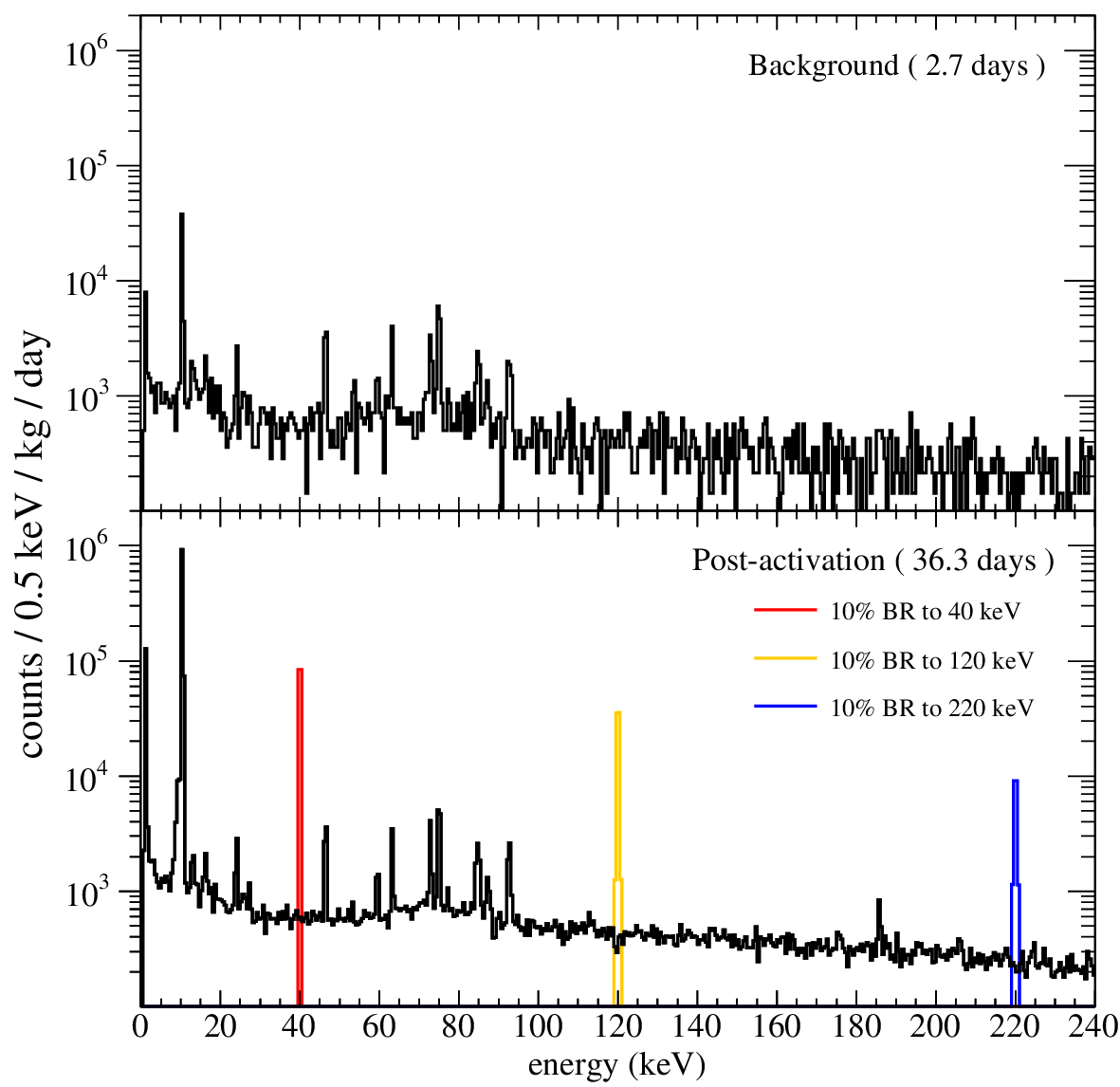}% Here is how to import EPS art
\caption{\label{fig:epsart} Pre- and post-activation spectra. Two enhanced peaks from $^{71}$Ge decay are the only significant difference. Example signatures (colored peaks) from decays to new excited $^{71}$Ga levels  with sufficient BR to relax the evidence for the gallium anomaly are shown. These are strongly disfavored  (see text).}
\end{figure}

The top panel in Fig.\ 1 shows the decay of the activity under the 1.29 keV and 10.37 keV peaks characteristic of $^{71}$Ge EC from the atomic L-shell and K-shell, respectively. These peaks can be observed in Fig.\ 2 as the only noticeable outcome from neutron exposure, in the spectral region measured. The $^{71}$Ge half-life derived from a fit to their summed rate is 11.46$\pm$0.04 d, in  excellent agreement with the 11.43$\pm$0.03 d assumed in the interpretation of experiments responsible for the gallium anomaly \cite{giunti,kopp,haxton}. This measured half-life is robust against the  procedure employed to extract the rates in Fig.\ 1. 

Low-background searches for rare processes involving large-mass germanium diodes can measure this half-life. However, the modest decay rates typically observed lead to much larger statistical uncertainties, as the activation of these detectors is only due to low-flux environmental neutrons during detector construction. The two peaks of interest here can also be contaminated by a longer-lived cosmogenic activation in $^{68}$Ge \cite{ppc}. Most importantly, the activation of other radioactive species in these larger crystals and in their cryostats can result in deviations from the expected half-life, for some modes of data treatment. Still, some of these unpublished $^{71}$Ge half-life measurements available to us are worth mentioning, as  all support the accepted value: 10.43$\pm$0.30 d, 10.91$\pm$0.91 d, 11.57$\pm$2.66 d, for detectors in \cite{cogent}, \cite{cosme}, \cite{sg}, respectively. Of special mention is the  intense accidental activation of a 440 cm$^{3}$ p-type germanium crystal \cite{ppc}. High-statistics data from this detector (Fig.\ 1, lower panel) point to an 11.80$\pm$0.05 d half-life. This value should be considered less reliable than that from our {\it ad hoc} measurement, for the reasons above and the shorter time span involved.

Fig.\ 2 shows the pre- and post-activation spectra in the present measurement. All peak-like features in the second appear in the first, i.e., we find no evidence for a non-negligible BR to new short-lived excited states of $^{71}$Ga. Three colored peaks superimposed on the post-activation spectrum show the expected magnitude of signals from de-excitation gammas generated by such a phenomenon, for a 10\% BR capable of relaxing the gallium anomaly to a $\sim 3  ~\sigma$ statistical evidence \cite{kopp}. Those include the effect of energy resolution and simulated efficiency for full-energy detection of gammas internally emitted in the detector. The most significant peak-like structure in this spectrum not observed pre-activation corresponds to a mere 0.4\% BR, which has negligible impact on the gallium anomaly \cite{kopp}. This other possible path for its resolution is therefore not supported by our data, with the  caveat that any new excited level(s) might be sufficiently long-lived (T$_{1/2}\gtrsim$ 12.6 yr) to escape our  0.4\% BR sensitivity. 

A final property of $^{71}$Ge EC decay able to impact the interpretation of the gallium anomaly, not considered in \cite{giunti,kopp}, are the relative EC rates from different atomic shells. Similarly to half-life and BR to the $^{71}$Ga ground state, those rates appear explicitly in the calculation of  $\sigma (\nu_{e} + ^{71}$Ga $\rightarrow$ e$^{-} + ^{71}$Ge) \cite{haxton}. The value of P$_{L}$/P$_{K}$ = 0.117 used in \cite{haxton} is traceable to proportional-counter studies dating back to 1971 \cite{LK}. Present data allow to measure this ratio with a different (and arguably more straightforward) technique, from the relative intensity of  1.29 keV and 10.37 keV peaks (Fig.\ 1, insets). We find P$_{L}$/P$_{K}$ = 0.116$\pm$0.004 for the data from \cite{ppc}. Following \cite{haxton}, the slightly larger  0.125$\pm$0.008 from the present detector would result in a reduction in $\sigma (\nu_{e} + ^{71}$Ga $\rightarrow$ e$^{-} + ^{71}$Ge) by less than 1\%.  Both measurements are in good agreement with a recent theoretical value of 0.12258(17) \cite{xavier,xavier2,xavier3}. We notice that P$_{M}$/P$_{L}$, beyond the reach of our detectors but also entering the derivation of the cross section, was recently measured at 0.16$\pm$0.03 \cite{cdms}. This is again in good agreement with the value of 0.165 adopted in \cite{haxton}.

In conclusion, our data strongly constrain any  explanation for the gallium anomaly based on the decay of $^{71}$Ge. As far as this specific input is concerned, the statistical significance  of the anomaly remains as large as $ 6 ~\sigma$ in some analyses (e.g., Fig.\ 1 in \cite{kopp}).

We are indebted to Wick Haxton, Joachim Kopp and Xavier Mougeot for useful comments.

\bibliography{apssamp}% Produces the bibliography via BibTeX.

\end{document}